\documentclass[12pt,preprint]{aastex}

\def\lya{\ifmmode {\rm\,Ly\alpha}\else
                 ${\rm\,Ly\alpha}$\fi}
                 
\begin{document}

\title{GRAPES, Grism Spectroscopy of the Hubble Ultra Deep Field: Description and Data Reduction}
\author{N. Pirzkal\altaffilmark{1}, C. Xu\altaffilmark{1}, S. Malhotra\altaffilmark{1}, J. E. Rhoads\altaffilmark{1}, A. M. Koekemoer\altaffilmark{1}, L. A. Moustakas\altaffilmark{1}, J. R., Walsh\altaffilmark{9}, R. A. Windhorst\altaffilmark{3}, E. Daddi\altaffilmark{9}, A. Cimatti\altaffilmark{4}, H. C. Ferguson\altaffilmark{1}, Jonathan P. Gardner\altaffilmark{6}, C. Gronwall\altaffilmark{2}, Z. Haiman\altaffilmark{5}, M. K{\"u}mmel\altaffilmark{9}, N. Panagia\altaffilmark{1}, A. Pasquali\altaffilmark{8}, M. Stiavelli\altaffilmark{1}, S. di Serego Alighieri\altaffilmark{4}, Z. Tsvetanov\altaffilmark{10}$^,$\altaffilmark{11}, J. Vernet\altaffilmark{4}, H. Yan\altaffilmark{7}}

\altaffiltext{1}{Space Telescope Science Institute, 3700 SanMartin Drive, Baltimore, MD21218, USA}
\altaffiltext{2}{Department of Astronomy \& Astrophysics, Pennsylvania State University, 525 Davey Laboratory, University Park, PA 16802}
\altaffiltext{3}{Dept. of Physics \& Astronomy, Arizona State University, Street: Tyler Mall PSF-470, P.O. Box 871504, Tempe, AZ 85287-1504, USA}
\altaffiltext{4}{INAF- Osservatorio Astrofisico di Arcetri, Largo E. Fermi, 5. I-50125 Firenze, Italia}
\altaffiltext{5}{Department of Astronomy, Columbia University, 1328 Pupin Hall, 550 West 120th St., New York, NY, 10027, USA}
\altaffiltext{6}{Laboratory for Astronomy and Solar Physics, Code 681, Goddard Space Flight Center, Greenbelt MD 20771}
\altaffiltext{7}{Spitzer Science Center, California Institute of Technology, Mail-Stop 100-22, Pasadena, CA 91125}
\altaffiltext{8}{Institute of Astronomy, ETH H\"{o}nggerberg, 8093 Zurich, Switzerland}
\altaffiltext{9}{ESO/ST-ECF, Karl-Schwarschild-Strasse 2, D-85748, Garching bei M\"{u}nchen, Germany}
\altaffiltext{10}{Department of Physics and Astronomy, Johns Hopkins Unuversity, 3701 San Martin Dr., Baltimore, MD 21218}
\altaffiltext{11}{NASA Headquarters, 300 E Str., SW, Washington, DC 20546}

\begin{abstract}
We present deep unbiased spectroscopy of the Hubble Ultra Deep Field (UDF) 
carried out using the slitless grism spectroscopy mode of the Advance Camera for Surveys on the Hubble Space Telescope (HST). The GRIsm ACS Program for Extragalactic Science (GRAPES) achieves continuum detection as faint as $z_{AB}=27.2$ using 40 orbits ($9.2 \times 10^4$ seconds) on HST. 
The data were taken at four orientation angles to correct for the overlap of spectra.  GRAPES data provide a unique, uninterrupted, low resolution (R=100) spectral coverage for $5500\AA < \lambda < 10500\AA$, and allow us to detect high redshift galaxies at  $4 < z < 7$ whether they have $\lya$ lines or just show the Lyman Break, as well as find low luminosity AGNs  in an unbiased fashion. This paper describes in detail the observations and the data reduction, and  examines the quality of the extracted spectra. Subsequent papers will deal with the analysis of the data. The extracted and calibrated GRAPES spectra will be available from MAST at STScI.
\end{abstract}

\section{Motivation}

The most common route for identifying the redshift and nature of distant
extragalactic objects involves selection on the basis of broad-band color
(e.g. Lyman Break Galaxies: \citet{Steidel2003}, quasars: \citet{Fan2003}),
narrow-band excess (\citet{Rhoads2001}, \citet{Ouchi2003}, \cite{Hu2002}), or photometric redshift
\citep{Connolly1997} followed by spectroscopic follow-up.

With the grism on ACS we are able to target all objects in the Hubble Ultra Deep Field in an unbiased way. For compact objects
(Semi-major axis a=0.25'') typical of high redshift galaxies we achieve continuum
detection down to  an AB magnitude of $z_{AB}=27.2$ (ACS F850LP filter). The lack of OH emission bands in space,
Hubble's superb resolution and red-sensitive CCDs on ACS all combine
to make this one of the deepest spectroscopic samples available. While the grism spectra
are low resolution, they
are adequate to spectroscopically confirm galaxies between
redshifts 4 -- 7 by their Lyman-break at 1216 \AA, to detect  prominent emission 
lines from both normal galaxies and AGNs, and to detect the 
4000 \AA\ break at redshifts between 0.5 -- 1.5.

\section{GRAPES}
\begin{figure}[h]
\plotone{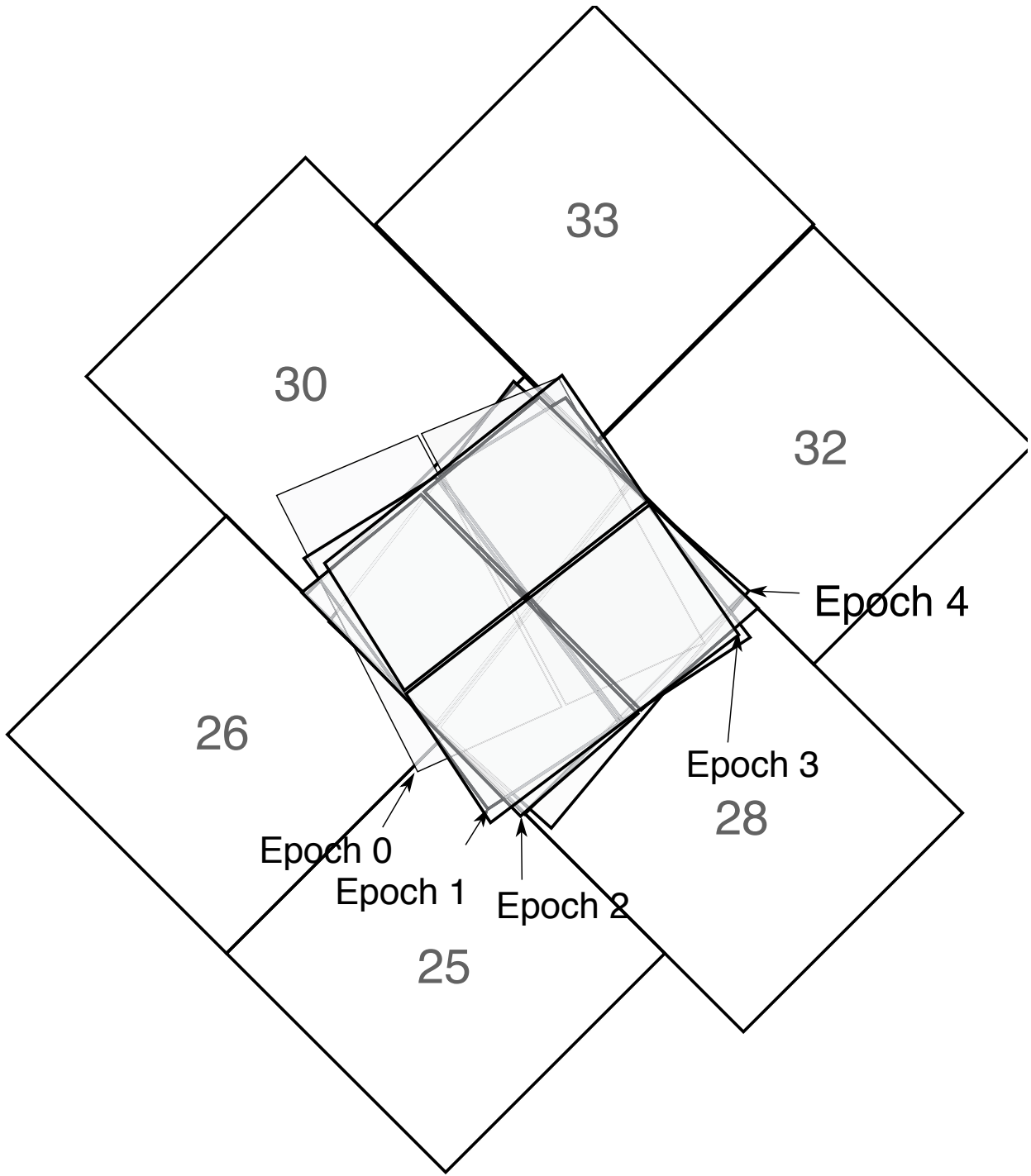}
\caption{The positions of the 5 grism observation epochs discussed in this paper plotted on top of the GOODS ACS  tiles. The GRAPES field matches the position of the HST Ultra Deep Field and roughly to Tile 23 of the GOODS field and ACS GOODS observation 29. North is up, East is left  on this figure.\label{orient}}
\end{figure}

\begin{figure}[h]
\plotone{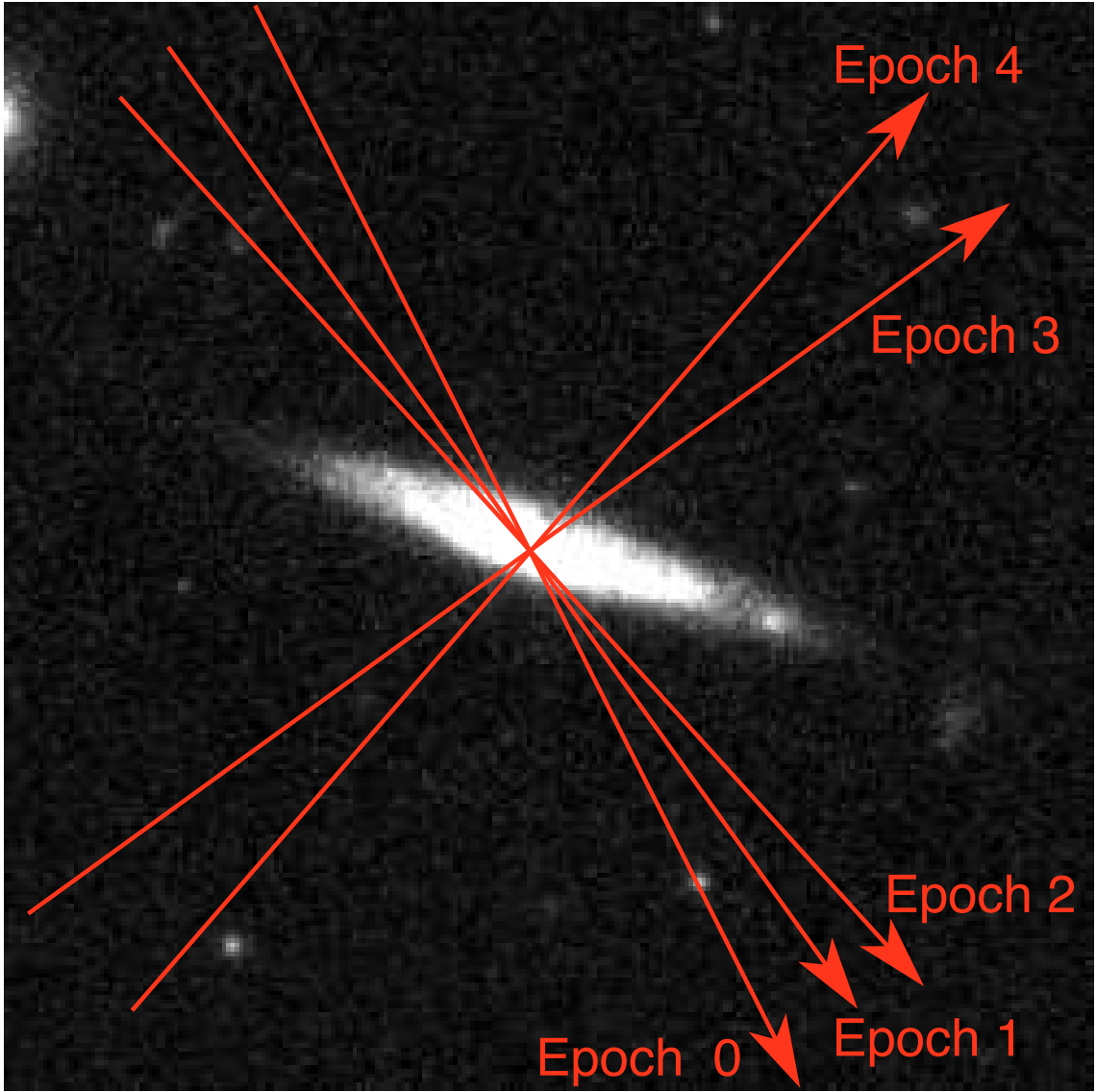}
\caption{The 5 orientations used to observe the GRAPES field. The arrows represent the grism dispersion direction in each of the GRAPES epochs. The labels near the arrows refer to the GRAPES epoch 0 through 4 listed in Table \ref{table1}. North is up, East is left on this figure\label{orient2}}
\end{figure}

GRAPES consists of 40 HST orbits of pointed ACS grism (G800L) observations of the Ultra Deep Field (UDF), taken during the HST Cycle 12. The ACS G800L mode is a slitless spectroscopic mode sensitive to wavelengths ranging from about $5500\AA$ to $10500\AA$, with a resolution R=100 (40 \AA\ per pixel resolution and a point source FWHM around 1.5 pixel). The GRAPES field overlaps mostly Tile 23 of the GOODS-S field. This is a field for which large amounts of multi-wavelength data already exist and for which deep spectra and spectroscopic identification of a large fraction of the sources can greatly improve the utilization of these data. The UDF was observed using the ACS grism for a total of  $9.2 \times 10^4$ seconds. The observations were taken at different epochs and at four different orientations (Position angles (PA\_V3) of 126, 134, 217, and 231 degrees, see Figure \ref{orient}. These angles refer to the HST V3 axis positions). The UDF was observed using these orientations in order to mitigate the effects of overlapping grism spectra in our slitless grism data. The relative angular differences between the GRAPES epochs were chosen after simulations showed that these could increase the number of spectra which remained uncontaminated by other spectra in at least one of the GRAPES epoch.  Using several orientations strongly increases the possibility that the spectrum of a given object would be uncontaminated in at least one of the observations. The pointing, observing dates, and integration times of each of the four GRAPES epochs are summarized in Table \ref{table1}.   The spectral dispersion directions at each of the five epochs are shown in Figure \ref{orient2}. Each individual epoch listed in Table \ref{table1}  consists of a small number of direct images taken using the F606W broad band ACS filter, and of a much larger number of G800L observations (see Table \ref{table1} for details). While the direct broad band imaging was relatively shallow, it was designed to be deep enough to astrometrically register our observations with the much deeper GOODS-S and UDF observations.\\
 
In addition to the four main  GRAPES epochs discussed above, we also made use of pre-existing ACS grism observations of the UDF field taken almost one year prior to our first epoch (PropID: 9352, PI: Riess, \citet{Riess2004}). These additional grism observations (epoch 0 in Table \ref{table1}) have a comparable depth to each of our GRAPES epochs, are only marginally offset from the other GRAPES observations, and were taken at a different position angle than the GRAPES observations (PA\_V3=117 degrees). Using these publicly available observations allowed us to increase the cumulative grism exposure time of the GRAPES observations by nearly one fifth (see Table \ref{table1}), bringing the total grism exposure time for the data discussed in this paper to $1.1 \times 10^5$ seconds.

\section{Peculiarities of ACS grism spectra}
The GRAPES grism spectra are slitless which results in certain peculiarities which need to be considered when extracting and using  these spectra: Every object in the field (and objects outside of the field whose spectra get dispersed within the field of view) produce a spectrum. First, the spectra are low resolution: $60 \AA$ nominally for point sources. Second, the background sky level is higher than in slit spectroscopy as each pixel sees the dispersed contribution at many wavelengths from different parts of the sky. Scaling and subtracting a high signal to noise super-sky estimate is required in order to bring the background residual errors down in the spectra of faint objects (Section \ref{backsub}). Third, the spectra of objects are all dispersed in one direction and can (and often do) overlap. To mitigate this problem we took data in different orientations to vary the dispersion direction relative to the objects positions in the field, and hence varying the amount of spectral overlap (see Section \ref{contam} where we discuss estimating the amount of spectral contamination in each spectrum). 
Finally, large sources have additional problems: The effective resolution of the spectrum of an extended object is degraded just as a wider slit degrades the spectral resolution in slit spectra.  The spatial extent of a large object can also cause the blue and red ends of a slitless, flux calibrated spectrum to roll up and wrongly appear to contain large amount of flux. This is because the observed spectrum is the intrinsic spectrum of the object convolved with its spatial structure (wavelength dependent) in the dispersed direction. This effect can be modeled, if some assumptions are made about the object shape as a function of wavelength, but cannot be generally corrected.

\section{Data Reduction and object lists\label{dataredux}}

The GRAPES data were obtained in a way that made them readily useable with aXe \footnote{\anchor{http://www.stecf.org/software/aXe/}{http://www.stecf.org/software/aXe/}}, the slitless spectroscopy extraction package developed at ST-ECF. This package is designed to perform batch extraction of thousands of spectra from individual ACS G800L observations \citep{Pirzkal2001b}. This extraction package took care of most of the low level calibration aspect of the extraction process. The ACS grism mode field dependence (the variation of the grism resolution as well as the variation of the tilt of the dispersed spectra on the detector as a function of position on the detector), the sensitivity of the G800L grism, and the pixel to pixel wavelength dependent flat-field of the ACS Wide Field Camera (WFC) are well calibrated and properly taken into account by the aXe software. 

The systematic and random error in the wavelength calibration is estimated to be $20\AA$ \citep{Pasquali2003}. The ACS wavelength dependent flat-field is accurate to within about $1\%$ at  wavelengths smaller than $9000\AA$, and to about $2-3\%$ at the redder wavelengths where fringing might affect the flat fielding. The error in the absolute flux calibration (not accounting for any aperture correction) is  $3\%$ from $5000\AA$ to $9000\AA$ and $20\%$ past $10000\AA$ \citep{Walsh2004}.

aXe produces a FITS file containing multiple FITS extensions, each containing a wavelength and flux calibrated extracted spectrum.
In order to extract spectra from a G800L image, aXe requires a list of objects believed to be present in the field. However, one needs to bear in mind that the ACS WFC suffers from a significant amount of distortion (the field of view, when projected onto the sky is significantly skewed in the corners). This effect is routinely dealt with by the ACS Data Pipeline in the case of direct imaging data, but with grism data this distortion correction cannot be performed by the data reduction pipeline. The extraction of spectra must therefore be performed on distorted grism images where the positions of the spectra are skewed with respect to the true position of the object on the sky. Our method for properly generating an object list for each GRAPES grism observation is described in more detail below.\\

As a first step in the data reduction process, all the data, direct and grism, were processed using the version 4.1 of CALACS \citep{CALACS} using the Best Reference Files from the STScI Archive. This step was handled automatically for us by the HST Data Archive. The direct images were processed using normal bias, dark, flat-fielding, and gain corrections using CALACS. For reasons mentioned above, the HST Archive final products, which are drizzled and geometrically corrected, were discarded. We instead used only the individually processed, non combined ACS datasets (FLT products). The grism observations were similarly processed using the CALACS based HST Data Pipeline.  The flat-field used by CALACS for the G800L data was a unity flat so that no pixel to pixel corrections were applied at that time, but so that quadrant gain correction would be properly applied (CALACS applies the flat-field and the gain correction at the same time). In the end, the G800L data were therefore bias, dark, and gain corrected, but no pixel to pixel flat field correction was applied. The flat-fielding of the grism data was handled by the extraction software aXe during the extraction process as explained in more details below.\\

\begin{deluxetable}{cccccll}
\tablecaption{Summary of the GRAPES Observations. Epochs 1-4 are the main GRAPES epochs. Epoch 0 consists of previously observed data which were obtained from the HST  Data Archive. The number of individual exposures is indicated in square brackets next to the exposure time. (*) Epoch zero direct images were taken using the F850LP filter, all other epochs used the F606W filter. \label{table1}}
\tablehead{\colhead{Epoch} & \colhead{PA\_V3} & \colhead{Observing} & \colhead{RA} & \colhead{Dec} & \colhead{Direct (F606W)} & \colhead{Grism (G800L)} \\
 & & Date & & &  \colhead{Seconds} &  \colhead{Seconds}
}
\startdata
0 & 117 & 2002 01 10& 3:32:37.52 & -27:46:46.55 & 2000.0 [4] (*)& 18840.0[16] \\
1 & 126 & 2003 22 09& 3:32:39.00 & -27:47:29.10 & 708.0 [2] & 24667.0 [17]\\
2 & 134 & 2003 01 10& 3:32:39.00 & -27:47:29.10 & 708.0 [2] & 24667.0 [17]\\
3 & 217 & 2003 02 12& 3:32:39.00 & -27:47:29.10 & 708.0 [2] & 21644.0 [15]\\
4 & 231 & 2003 17 01& 3:32:39.00 & -27:47:29.10 & 708.0 [2] & 21644.0 [15]
\enddata
\end{deluxetable}

\subsection{Aligning images and cosmic ray removal}
Cosmic rays and other cosmetic defects, which are not removed by CALACS,  were identified and removed from both the direct and grism images using the IRAF task Multidrizzle \citep{Koekemoer2002}. Multidrizzle  is an implementation of the Blot/Drizzle technique \citep{Drizzle} which allows for bad pixels and cosmic rays to be identified in each image of a stack, whether these have been dithered or not. This method computes a high-signal to noise, combined, cosmic-ray free image of the field of view and then replaces bad pixels in individual images with values drawn from this cleaned up image. The result of processing data with Multidrizzle is that of 1) generating deep stacks of both the direct and grism images which have been geometrically corrected (to compensate for the large amount of distortion), and of 2) generating copies of the original data which have been completely cleaned of CRs and other bad pixels, but are otherwise not modified in any way. In order to properly remove cosmic rays from our images, Multidrizzle requires a very good {\it a-priori} knowledge of the relative shifts and rotations between our images. We therefore first determined these carefully for both the set of shallow direct and the larger set of grism GRAPES images.
\\
The shallow direct images from GRAPES were affected by cosmic rays and only contained a few bright point sources. In order to align these images and removes the effects of the cosmic rays,  we started by selecting 10-30 compact sources to serve as a matching set. This matching set was used to compute initial offsets and rotations between each individual direct images. The images were individually drizzled and the positions of the matching sources were individually measured in the singly drizzled images using 3 different algorithm (centroiding, centroiding with masking nearby sources,  2-D IRAF gaussfit). The set of 3 independently measured positions were averaged if they agreed to within 0.3 pixel. If they did not agree, the corresponding source was removed from the matching set  and the shifts and rotations between images were recomputed. This method had two advantages: 1) some "well behaved" extended sources could be used to align our shallow direct imaging GRAPES images since these did not contain enough bright point sources, 2) cosmic-ray contaminated sources could be effectively removed from the source list (note, the singly drizzled images are heavily contaminated by cosmic rays). The determined coordinates then are fed into a minimum $\chi^2$ fit algorithms to find the relative shifts and rotations. The residuals of this fit was within 0.1 pixel in all cases.
 \\
When determining the shifts and rotations between individual grism observations, we used the same method used with the GRAPES direct images but using the measured positions of 10-20 of the slightly dispersed (over 2-3 pixels) point-like zeroth order spectra (which are present in the ACS G800L grism images). The residual of the fit was within 0.2 pixel.
\\
\subsection{Generating the object catalog}
The grism component of the GRAPES data reaches significantly deeper than the GRAPES direct imaging.  We used the GRAPES direct images to astrometrically lock down the position of objects which were independently detected from a much deeper GOODS z band. The deeper GOODS z band image was assembled using publicly available data and in such a way that it matched the GRAPES field. This image was $5'25'' \times 5'25''$, which is larger than the individual GRAPES observations ($3'4'' \times 3'4''$).\\
As discussed above, aXe requires a catalog of objects which has been derived using a non-geometrically corrected direct image of the field. The position of objects fed to aXe should be given as if these were positions within a non-geometrically corrected direct image of the same field just prior to placing the G800L dispersive element in the optical path. The GOODS z band Public Data 1.0 image, on the other hand, is a completely reduced, geometrically corrected image of the field.
\\
We used Sextractor to generate a master catalog of the objects in the GRAPES field using the very deep GOODS broad band images.  We used a detection and analysis thresholds (DETECT\_THRESH,  ANALYSIS\_THRESH respectively) of 1.8 sigma, a  minimum number of pixels above threshold (DETECT\_MINAREA)  of 5 pixels, and used a 2 pixel FWHM Gaussian as a detection  filter (FILTER\_NAME). The catalog was manually edited to remove spurious detection near the edges of the GOODS mosaic. \\
At each of the five epochs listed in Table \ref{table1} we then proceeded as follows.  A few hundred objects between the master catalogue and the catalogue from the GRAPES shallow direct images were matched automatically and a minimum $\chi^2$ fitting (with 5 sigma clipping) was performed to map between the two sets of the coordinates. The master object catalog was then transferred to the field of view of each individual GRAPES epoch. 
 We only allowed for a first order transformation without scale changes (e.g.  translation plus rotation)  because both the shallow GRAPES direct image stacks, and the deeper GOODS image (from which the master object catalog was generated) were geometrically corrected. The RMS of the fit at each epoch was typically on the order of 0.3 pixels. This small amount of residual is indicative of the high levels to which the ACS optical field distortions have been calibrated.
\\
Once the UDF object catalog became public and available to us, we revised our master object catalog to include all of the sources included in the UDF catalog, plus bright sources from the GOODS catalog which are outside of the GRAPES field of view but likely to contaminate the spectra of the objects in the GRAPES field. Restricting ourselves to sources with $z_{AB} < 29.5$ (which is conservatively deeper than what we expect to reach with the GRAPES spectroscopic observations), our final object master catalog contained 5138 UDF objects with $z_{AB}$ ranging from 14.9 to 29.5. Our master object catalog contains an additional 989 objects which were selected from our GOODS z band and the GOODS public catalog and which are within 10'' of the GRAPES field (200 WFC pixels). These we included solely to later estimate the amount of spectral contamination in the GRAPES calibrated spectra (see Section \ref{contam}).
\\
The object catalogs generated for each of the GRAPES epochs still required some processing in order to compute object catalogs that corresponded to individual G800L observations (i.e. FLT files). Because the G800L and F606W GRAPES observations were always  taken using the same guide stars and within the same visit at each epoch listed in Table \ref{table1}, we were able to use the World Coordinate System (WCS) of both the F606W  and G800L stacks to accurately compute the position of objects in the the G800L image stacks (better than 0.5 of a pixel, see \citet{Pirzkal2001}). 
These newly created object list, transformed to the reference frame of G800L image stacks, together with the knowledge of ACS geometric distortion, and of the offsets and rotations between individual images in the image stack, ultimately allowed us to accurately compute the position of each object in each of the individual GRAPES G800L observations. The IRAF task TRANBACK, was used to perform this last computation. This task uses exactly the same code as the IRAF DRIZZLE task does and therefore handles the issues of reference pixels, geometrical distortion, orientation, input offsets and rotations in a completely self consistent manner.  In the end, the series of manipulations we just described allowed us to individually produce complete (down to the levels reached by our master object catalog described above) lists of object for each of our G800L GRAPES observations, using only a relatively small amount of GRAPES direct imaging.

\section{aXe extraction}
Following the generation of individually cosmic ray cleaned, bias subtracted and gain corrected G800L datasets, and the computation of object positions in each of these G800L datasets, we used aXe to perform the extraction and calibration of the extracted spectra. 
\\
\subsection{Background estimate}\label{backsub}
An improper estimate of the local background of a faint spectrum can lead to large errors in both the flux calibration of the spectrum as well as its shape. The level of the background is expected to vary slightly from one observation to the next (it is a combination of zodiacal light and earth limb reflection). aXe was designed to allow one to compute an estimate of the background of an observation by masking out the locations of known spectra and replacing these by interpolating the levels of neighboring regions. This method was found to work relatively well for bright objects but not well enough for faint spectra because it does not account for small smooth, correlated  variation of the background as a function of position on the detector. We therefore chose to use a G800L mode super-sky estimate which we generated using 84 individual G800L observations obtained from the HST Archive. These datasets were completely independent of the data obtained for this project. Spectra in these datasets were located using Sextractor \citep{Sextractor} by simply finding groups of connected pixels one or more sigma above the local background. The pixels thus detected in each dataset were marked in an associated IRAF BPM  bad pixel map. This BPM map, containing the set of pixels we want to avoid when using these images with IRAF tasks, was enlarged slightly by convolving it with a 20 pixel wide gaussian. Each dataset was  then normalized by its mode using a 10 iteration, 3 sigma rejection algorithm, and the resulting 84 masked and scaled datasets were finally averaged together using the IRAF task imcombine.  A high signal to noise estimate of what an observation of an empty part of the sky would look like when observed using the ACS WFC G800L grism mode was thus generated. We further checked and improved this super-sky by scaling and subtracting our super-sky from each of the initial 84 datasets. Four datasets were found to have too few unmasked pixels and to be a somewhat poorer fit to our super-sky estimate. These were removed from our list and the remaining 80 datasets were used once again to compute a slightly improved version of our super-sky. It is this later iteration of the super-sky which was used  for the spectral extraction of the GRAPES data described in this paper. \\

\subsection{Extraction}\label{extraction}
We extracted individual spectra using the aXe software. First, we removed the background from each individual exposure. This was done by computing the scale factor between our super-sky and regions of the observation which was a-priori known to not contain any spectral information (based on the much deeper object catalog we were using). We actually made use of the aXe Background Estimate (aXe\_BE) task  to determine which part of each individual observation should be used to compute this scale factor. Using the aXe task produced a more conservative selection of the background portion of the observation than when using SeXtractor as we did previously. Once the scale factor was computed, we scaled and subtracted our super-sky to the individual observation. This method worked well and proved to be robust. We estimate that the residual error level in our background subtracted observations to be within a few  $10^{-4}$ counts/s/pixel. \\
In addition to the improved background subtraction, we also made several conservative choices of aXe extraction parameters: while aXe does allow for the extraction of an object to not be perpendicular to the spectral trace (allowing for a highly elongated object to be extracted along the direction of its semi-major axis), we conservatively turned this feature off for objects with a semi-major axis size smaller than 10 pixels and with a measured ellipticity smaller than 0.5 to avoid the possibility of adding artifacts to the extracted spectra. We performed both narrow and wider extractions of the spectra. The extraction widths were set to twice the distance separating the tangent lines which are parallel to the dispersion direction (nearly horizontal on the detector) above and below the ellipse described by the Sextractor parameters A\_IMAGE,B\_IMAGE, and THETA\_IMAGE. The smallest allowed extraction width was set to 6 pixels. These parameters were chosen to maximize the signal-to-noise in faint point sources and slightly larger objects at the expense of flux conservation.
\\
\subsection{Spectra combination}
Following their extraction using aXe, individual spectra of objects observed at the same epoch (and hence subjected to the same contamination from nearby objects and also with objects being similarly oriented with respect to the spectral dispersion direction) were combined together to improve their signal to noise. The combination was done by first  linearly interpolating each spectrum onto a common wavelength scale, and then by performing a 3 sigma rejection weighted (by exposure time) average at each wavelength (The 3 sigma rejection was used to ensure that we removed the effect of any left over cosmic ray or cosmetic glitch present in the individually extracted spectra). The earlier use of Multidrizzle to cleanup the original grism image made the 3 sigma rejection step largely redundant as the effect of most comic rays and bad pixels were almost always properly removed at the early stage of the data reduction process. 

\section{Contamination estimate\label{contam}}
The effect of the contamination from nearby neighbors is different from one epoch (i.e. orientation) to the next.  We recomputed the contamination level of each spectrum assuming that the object had a flat spectra (equal flux at each wavelength), using the sizes listed in the original Sextractor catalog, and the relative throughputs of the different ACS G800L dispersive orders. Using this information, and the position of every spectral order of every object in the field, we estimated the fraction of the flux at a chosen wavelength of a given spectrum which was contaminated by the flux originating from  neighboring spectra. This step was performed for each spectrum extracted at each epoch and the estimate of the contamination (in percent) was updated in the final version of the GRAPES extracted spectra.

\section{Availability of the reduced GRAPES spectra}
The extracted and calibrated spectra described in this paper are expected to be made available through the STScI MAST archive in June 2004. Information on how to retrieve these data can be obtained from http://www.stsci.edu/. The spectra are in the aXe SPC output product file formats and are multi-extension FITS files with each extension containing a binary table describing the spectrum. Users should refer to the aXe User Manual \footnote{Available from http://www.stecf.org/software/aXe/}, to get the detailed description of the content of these files. The contamination information in the SPC files was computed according to Section \ref{contam}. Spectra of objects observed at each of the epochs listed in Table \ref{table1} are being made available, including the computation of the spectral contamination which was described above. 

\section{Description of GRAPES spectra\label{Description}}
\begin{figure}[h]
\plotone{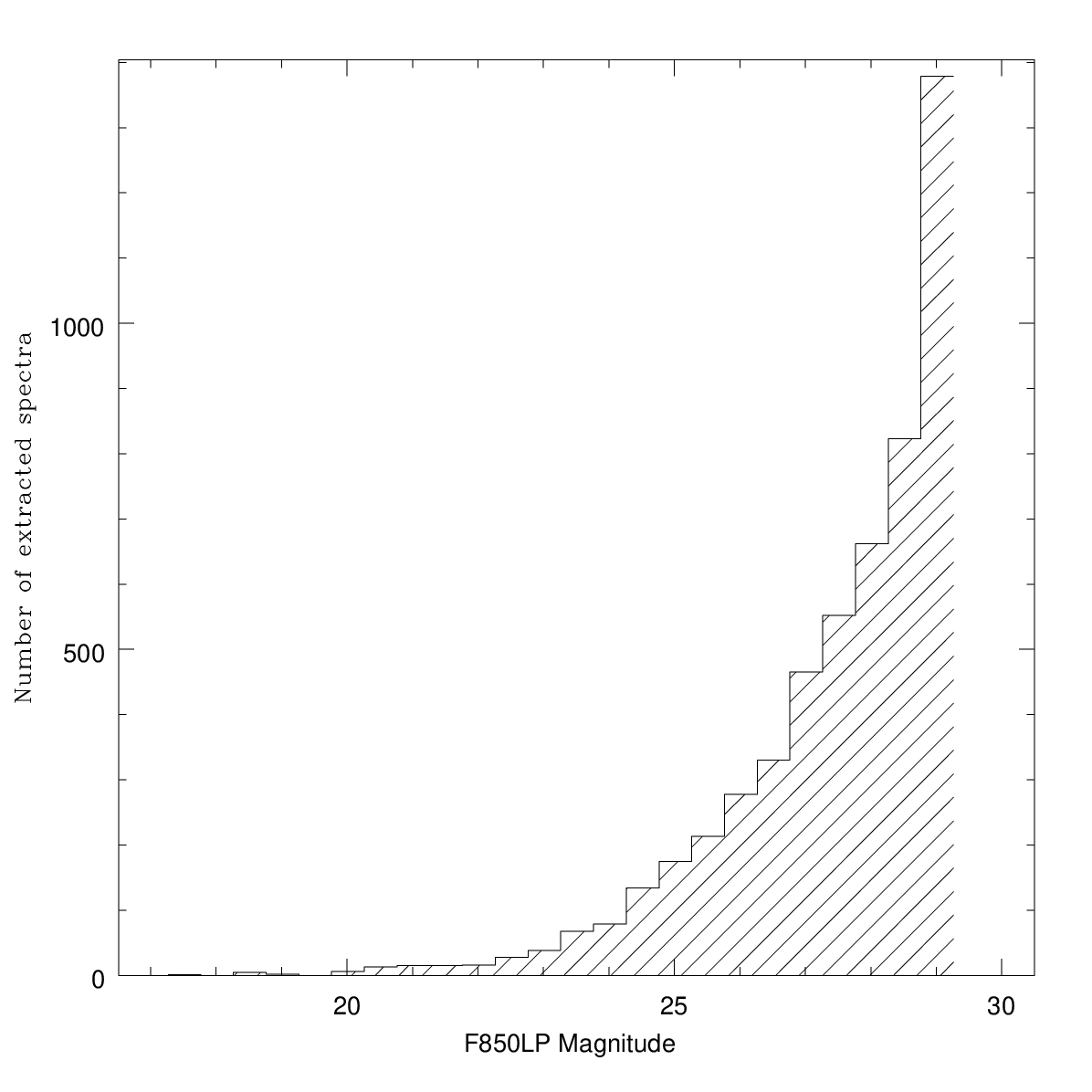}
\caption{Histogram of the $z_{AB}$ (F850LP) magnitude distribution of the GRAPES spectra. 1421 objects are brighter than $z_{AB} = 27$\label{maghist}}
\end{figure}

\begin{figure}[h]
\plotone{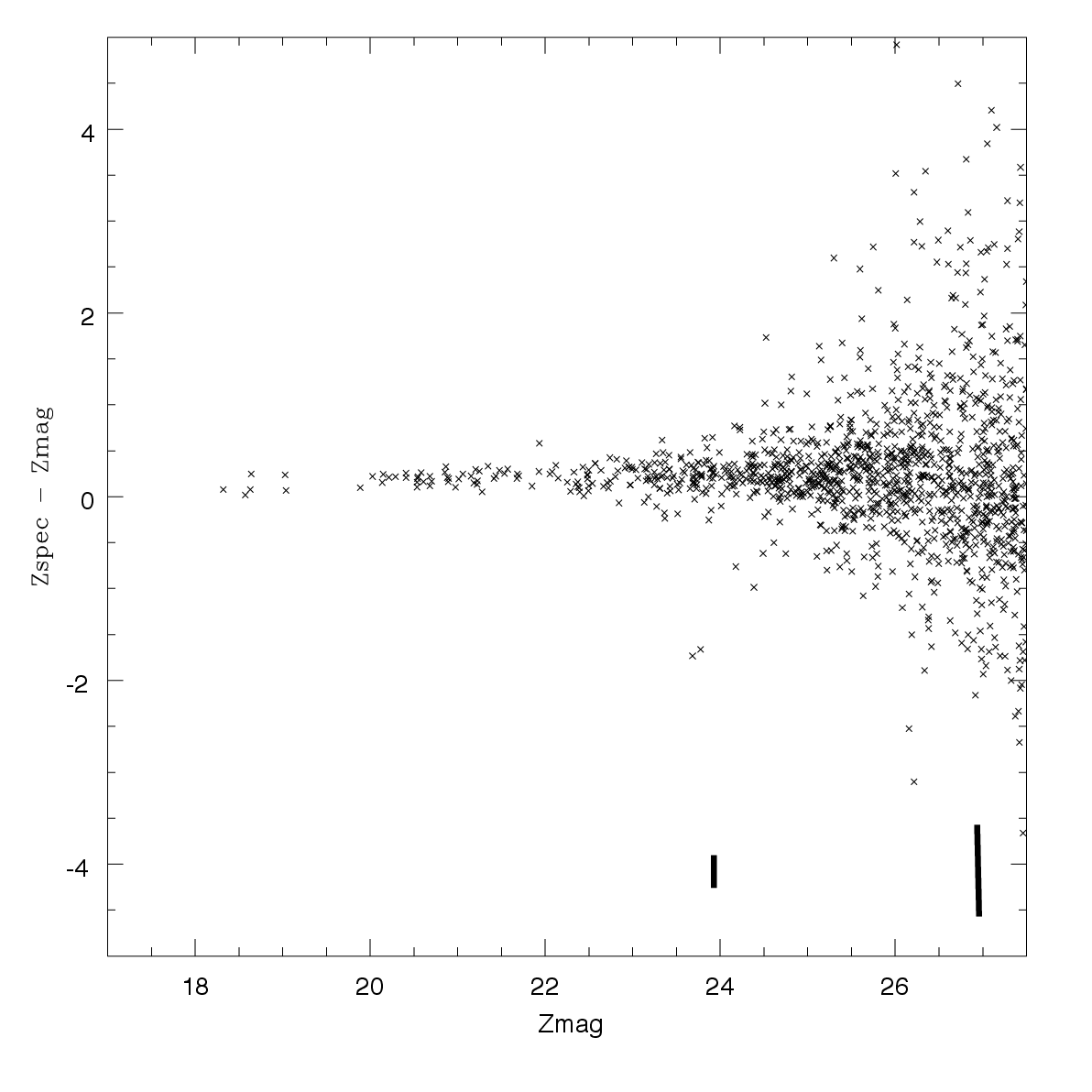}
\caption{Plot showing the difference in z band magnitudes measured from the direct UDF images (z-axis) and from the calibrated GRAPES spectra. The two agree to within the Poisson noise level. The relatively small turn off in the distribution at magnitudes fainter than 26.0 is indicative of an error in background subtraction that is smaller than a few $10^{-4}$ counts/s/pixel. The small black vertical lines at z=24 and z=27 are the $\pm1\sigma$ noise estimate of the synthetic magnitudes (which are noisier than the UDF published magnitude because they were derived from spectra which are spread over about 100 times more pixels than the ACS direct images.\label{dmag_vs_zmag}}
\end{figure}

\begin{figure}[h]
\plotone{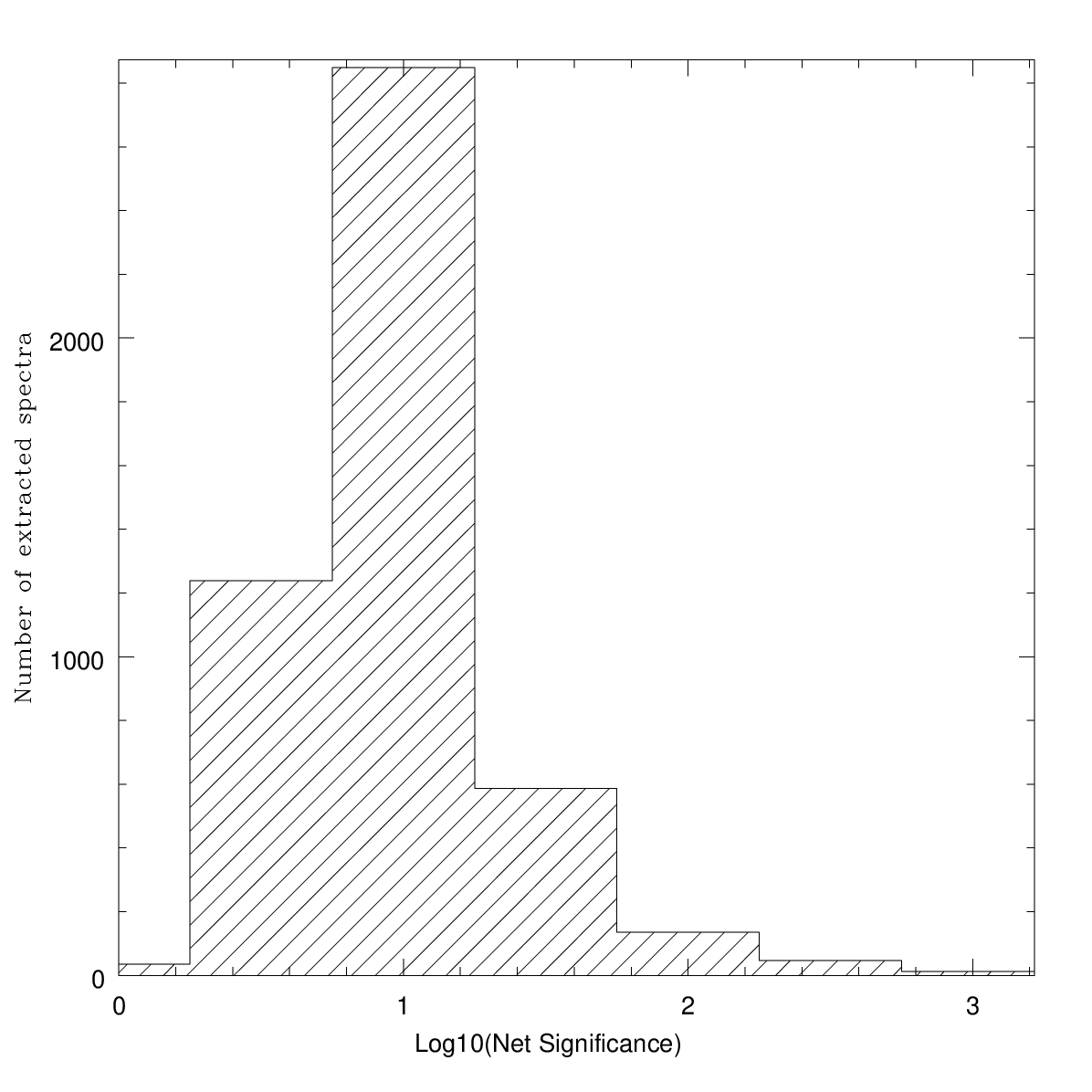}
\caption{Histogram of the net spectral significance {$\cal N$} (Defined as the maximum cumulative S/N of  a spectrum, see Section \ref{Description}) of the GRAPES epoch 1 spectra. 
{$\cal N$} generally correlates with the brightness of the source, except for sources with bright lines. Simulations show that {$\cal N$}=10 corresponds to a $5\sigma$ event in absence of any signal. 1680 spectra in the GRAPES epoch 1 have a  value of {$\cal N$} of 10 or above.
 \label{netsig}}
\end{figure}

\begin{figure}[h]
\plotone{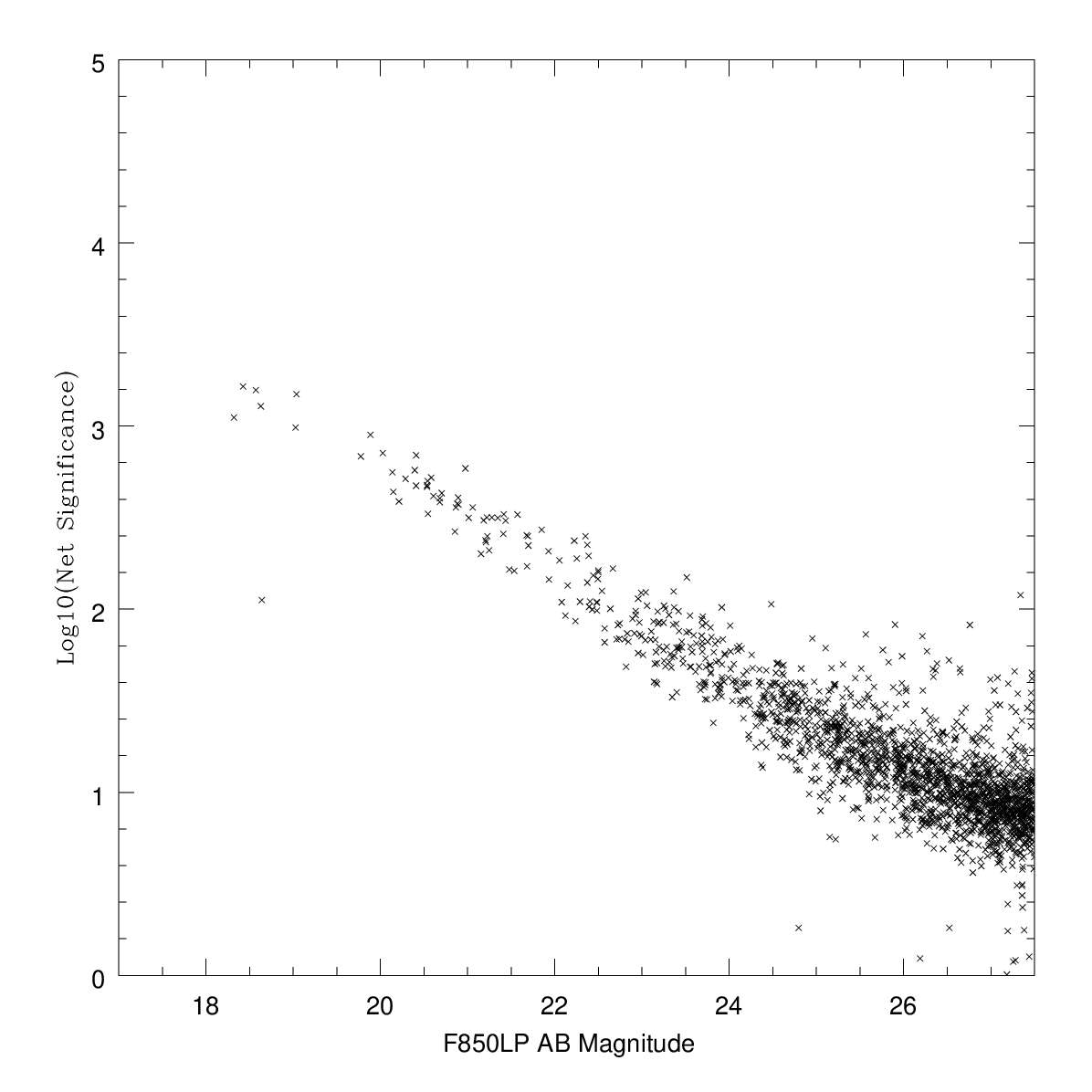}
\caption{Plot of the net significance (defined as the maximum cumulative S/N of  a spectrum, see Section \ref{Description}) of the GRAPES epoch 1 spectra vs. $z_{AB}$ (F850LP) magnitudes. The expected trend that as objects get dimmer their spectra get noisier, the measured net significance decreases on average. In this single epoch, a net significance of 10 corresponds on average to a magnitude $z_{AB}=26.2$. An object of $z_{AB}=27.1$ with 5 such spectra coadded would have a net significance {$\cal N$} of 10. \label{netsigmag}}
\end{figure}

5138 spectra were extracted and combined. Most of these spectra were observed at more than one epoch, but not all spectra remained completely uncontaminated by nearby objects at all epochs. The quality of the final combined spectra therefore vary somewhat from one spectrum to the next. 1421 objects are brighter than our nominal limit of $z_{AB}  =  27$ and potentially have useful spectra. A histogram of the magnitude of objects with extracted spectra is shown in Figure \ref{maghist}. The quality of the background subtraction process described in section \ref{extraction} is illustrated in Figure \ref{dmag_vs_zmag} where we plot synthetic z band magnitudes (obtained by convolving the GRAPES spectra by the ACS z band filter bandpass) to the published UDF z band magnitude of the same objects.  A way to access the information content of each spectrum is to compute its net significance ${\cal N}$, which we define as the maximum cumulative S/N of  a spectrum.  Figure \ref{netsig} shows the net significance distribution to peak below the value of 10 which is an indication that we attempted to extract extremely low S/N spectra (the correspondingly faint  objects were included in our object master catalog so that they would be accounted for when recomputing the level of spectral contamination as discussed in Section \ref{contam}) The significance was computed for each spectrum by sorting the resolution elements in increasing signal to noise, computing a running cumulative S/N using an increasing number of bins (of intrinsically decreasing S/N) up to the point where the cumulative S/N is observed to turn-over and adding additional data only ends up decreasing the resulting cumulative S/N.  A low significance implies that there is little information in the spectrum significantly above noise. A higher net significance implies that  either an isolated emission feature (e.g. $\lya$ emitter with little to no background), a significant level of continuum emission exists, or a combination of the two. Sorting pixels on signal to noise prior to calculating net significance
implies that even a pure noise spectrum will yield positive {$\cal N$}.
Simulations show that a one-dimensional pure Gaussian noise spectrum 
containing $n_{\rm pix}$ independent pixels and no true signal 
will have ${\cal N} \approx 6.35 \sqrt{(n_{\rm pix}/100)} \pm 0.72$.
Thus a spectrum with ${\cal N} \ > 8.5 \sqrt{(n_{\rm pix}/100)}$ corresponds
to a $3\sigma$ detection of real signal in the grism data. Out of the 5138 GRAPES spectra, 1680 spectra have a computed net significance greater than 10. This number is close to the number of objects brighter than $27^{th}$ magnitude and demonstrates the tight relation (on average) between an object magnitude and the quality of its spectrum (Figure \ref{netsigmag}). \\

\section{Sample spectra}
\begin{figure}[h]
\plotone{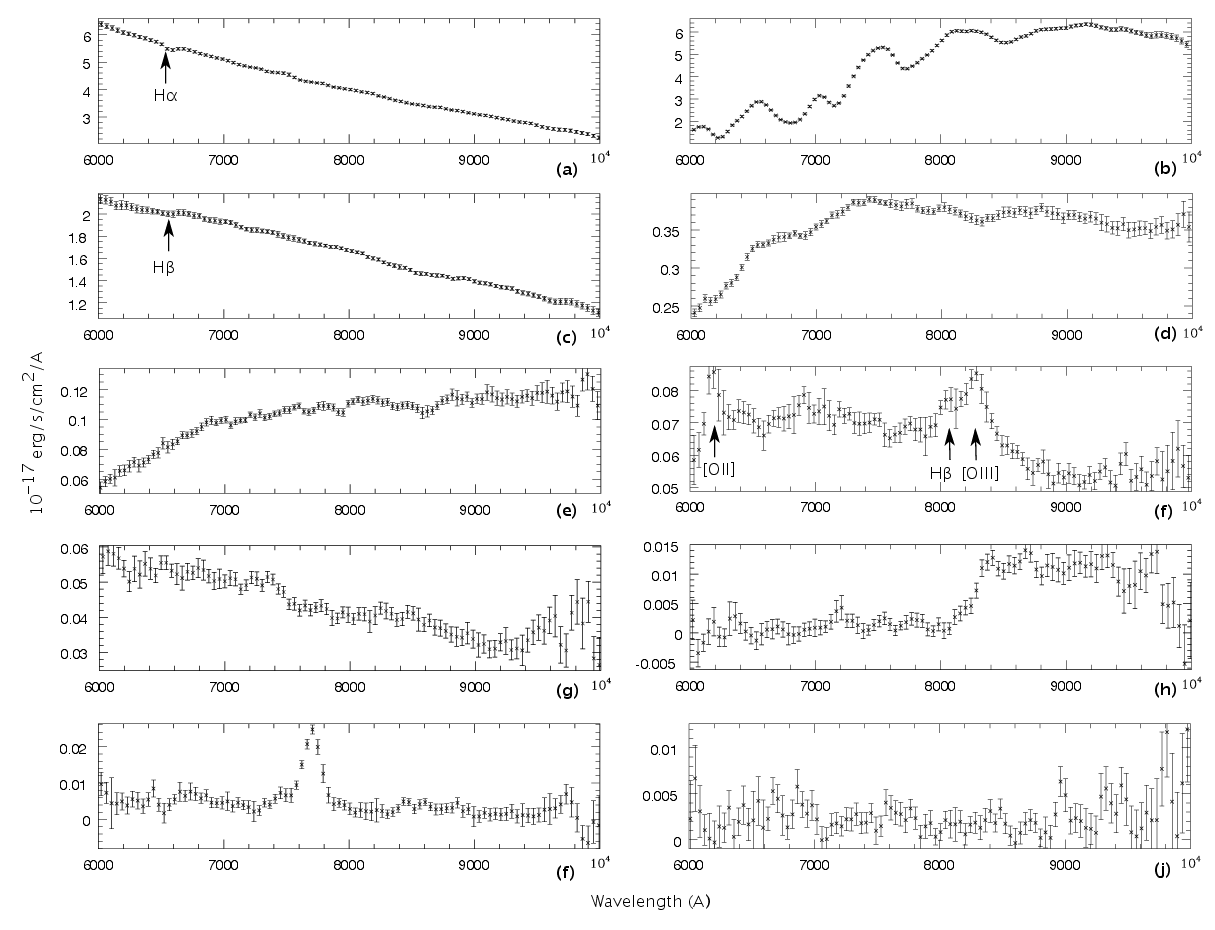}
\caption{Spectra of some objects with z magnitudes ranging from 18.3 to 27.1. The individual object $i_{AB}$ (F775W) magnitude, $z_{AB}$ (F850LP) magnitude and spectral net significance are listed in Table \ref{tablemultispec}. (a) and (c) show hot stars with $H\alpha$ absorption, (b) is a cool star, (d) is an elliptical galaxy at $z \sim 0.6$, (e) through (g) and (i) are star forming galaxies at intermediate redshift including object (f) at $z = 0.66$ which shows $H\beta$ as well as [OII] and [OIII] , (h) is a Lyman break galaxy at $z \approx 5.8$ and has been previously published (\citet{Stanway2003} and \citet{Dickinson2003}), and (j) is a faint continuum source near the detection limit of the GRAPES grism data.
\label{multispec}}
\end{figure}

\begin{figure}[h]
\includegraphics[width=5.0in]{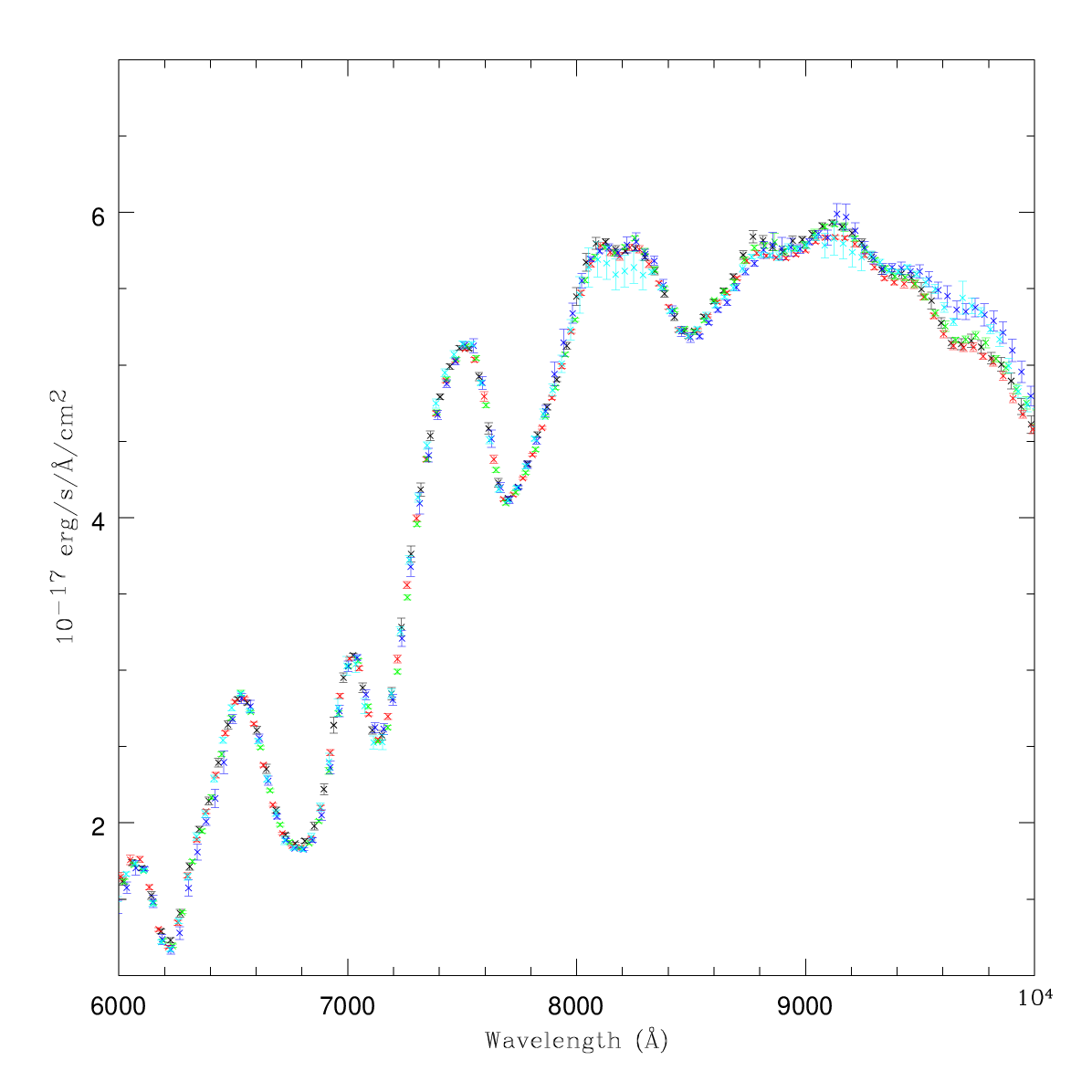}
\caption{The five individually extracted spectra of a dwarf star\label{multi6a}. The five extracted and calibrated spectra from the five GRAPES epochs listed in Table \ref{table1} are plotted in different colors on this plot. The wavelength and flux calibration agree very well up to 9500\AA where flat-fielding of the spectra becomes the main limiting factor in the flux calibration of the GRAPES spectra (see Section \ref{dataredux})}
\end{figure}

A few spectra from GRAPES are shown in Figure \ref{multispec} where we present spectra of 10 objects with z magnitude ranging from 18.3 to 27.1.  Table \ref{tablemultispec} summarizes the properties of these objects. These were obtained by averaging the 5 epoch observations available for each object. The spectra were simply resampled onto a common grid and averaged. No attempt was made to compensate for the differences in orientation of extended objects with respect to the dispersion direction. 
The quality of the flux and wavelength calibration is illustrated in Figure \ref{multi6a}, where we over-plotted the 5 individual epoch extracted spectra of a dwarf star, showing the agreement in wavelength and flux calibration between observations taken at several months intervals and using different parts of the ACS WFC detector.

\section{Conclusion}
We described the GRIsm ACS Program for Extragalactic Science (GRAPES) which is an ACS slitless spectroscopy survey of the Hubble Ultra Deep Field (UDF) using 40 orbits ($9.2 \times 10^4$ seconds) on HST.  We have extracted the spectra of 5138 objects, down to an AB magnitude in the ACS z band (F850LP) of 29.5, each observed at up to 5 distinct epochs and orientation on the field (1421 objects brighter than $z_{AB}=27$, 1680 objects have a net spectral significance which is above 10). These extracted data are to be made available via the STScI MAST data archive in June 2004. These spectra are low resolution slitless spectra ($40\AA$ per pixel, with a nominal point source resolution of $60\AA$) which are fully wavelength and flux calibrated. We estimate the absolute wavelength calibration (which is astrometry limited) to be within $20\AA$ and the absolute flux calibration to be within $3\%$ from $5000\AA$ to $9000\AA$. 

\acknowledgments
This work was supported by grant GO -09793.01-A from the Space Telescope Science Institute, which is operated by AURA under NASA contract NAS5-26555. This project has made use of the aXe extraction software, produced by ST-ECF, Garching, Germany.

\begin{deluxetable}{ccccc}
\tablecaption{The ID (from the UDF public catalog 1.0), $i_{AB}$ (F775W), $z_{AB}$ (F850LP), and computed spectral significance ({$\cal N$}) of the spectra shown in Figure \ref{multispec}\label{tablemultispec}}
\tablehead{\colhead{Object } & \colhead{ID} & \colhead{$i_{AB}$} & \colhead{$z_{AB}$} & \colhead{{$\cal N$}}\\
\colhead{in Figure \ref{multispec}} & \colhead{(UDF catalog)} & & & 
}
\startdata
a & 2150 & 19.13 & 18.32 & 1940.94 \\
b & 1147 & 19.17 & 19.04 & 1896.97 \\
c & 9230 &  20.00 & 19.89 & 1897.03 \\
d & 1960 &  21.53 & 21.23 & 1139.54 \\
e & 8576 &  22.30 & 22.39 & 353.77 \\
f & 5569 &  23.32 & 23.20 & 749.32 \\
g & 2927 &  23.45 & 23.84 & 207.24 \\
h & 2225 &  26.83 & 25.29 & 59.38 \\
i & 2241 &  25.86 & 26.22 & 35.62 \\
j & 5888 &  27.17 & 27.13 & 20.59 
\enddata
\end{deluxetable}

\end{document}